# Leveraging GNN to Enhance MEF Method in Predicting ENSO

Saghar Ganji[1], Ahmad Reza Labibzadeh[2], Alireza Hassani[3], Mohammad Naisipour[4]

1- B.Sc., Department of Computer Engineering, Shiraz University of Technology, Shiraz, Iran.
sagharganji82@gmail.com
2- MSC, Department of Computer, Sharif University of Technology, Tehran, Iran
a.r.labibzadeh@gmail.com
3- PhD student, Jacksonville University ,Jacksonville, Florida. ahassan4@jacksonville.edu
4- PhD, Department of Civil Engineering, Faculty of Engineering, University of Zanjan, Zanjan, Iran.
m.naisipour@znu.ac.ir

**ABSTRACT:**
Reliable long-lead forecasting of the El Niño–Southern Oscillation (ENSO) remains a long-standing challenge in climate science. The previously developed Multimodal ENSO Forecast (MEF) model uses 80 ensemble predictions by two independent deep learning modules: a 3D Convolutional Neural Network (3D-CNN) and a time-series module. In their approach, outputs of the two modules are combined using a weighting strategy wherein one is prioritized over the other as a function of global performance. Separate weighting or testing of individual ensemble members did not occur, however, which may have limited the model to optimize the use of high-performing but spread-out forecasts. In this study, we propose a better framework that employs graph-based analysis to directly model similarity between all 80 members of the ensemble. By constructing an undirected graph whose vertices are ensemble outputs and whose weights on edges measure similarity (via RMSE and correlation), we identify and cluster structurally similar and accurate predictions. From which we obtain an optimized subset of 20 members using community detection methods. The final prediction is then obtained by averaging this optimized subset. This method improves the forecast skill through noise removal and emphasis on ensemble coherence. Interestingly, our graph-based selection shows robust statistical characteristics among top performers, offering new ensemble behavior insights. In addition, we observe that while the GNN-based approach does not always outperform the baseline MEF under every scenario, it produces more stable and consistent outputs, particularly in compound long-lead situations. The approach is model-agnostic too, suggesting that it can be applied directly to other forecasting models with gargantuan ensemble outputs, such as statistical, physical, or hybrid models.

## 1 Introduction

Long-lead-time prediction of the El Niño–Southern Oscillation (ENSO) is one of the central problems of climate science since the system is inherently nonlinear, is dynamically chaotic, and involves complex teleconnections [4]. Successful prediction of ENSO is of crucial practical value due to its strong global impacts on the weather, agriculture, water resources, and socio-economic activity [33, 3]. Even though tremendous advances have been achieved in model innovations, it is still not reliable to predict the behavior of ENSO more than one year in advance, especially after 2000 when climate variability has been increasingly unpredictable [27, 95].

The advent of artificial intelligence has precipitated a transformative epoch, reshaping economic ecosystems while simultaneously redefining computational paradigms and capabilities [38].The recent breakthroughs in deep learning have introduced new techniques for nonlinear and spatio-temporal pattern extraction in ocean-atmosphere data[27, 24] . This is achieved in the Multimodal ENSO Forecasting (MEF) model, wherein two independent modules are 3D Convolutional

---

* Corresponding author: saeedpanah@znu.ac.ir

Neural Networks (3D-CNN) and a time-series forecasting module, both having 40 ensemble outputs. The original baseline MEF approach employed a weighted fusion framework, which combined the module outputs based on their aggregate uncertainties [27, 86]. However, this method did not blend individual strengths of ensemble members nor explore the inherent structure within the ensemble space. Consequently, extremely precise predictions were undermined in the process of averaging while less precise outputs could have been assigned undue influence [65].

To avoid this shortcoming, we present a GNN-augmented MEF methodology whereby not only the ensemble members' similarity is also expressed explicitly as a graph structure [25]. The ensemble runs are nodes, and edge weights are calculated based on measures of similarity such as RMSE and correlation [65]. With Graph Neural Networks (GNNs), we detect clusters of stable and accurate predictions from the ensemble. This enables us to choose a sub-set of the most structurally robust and statistically improved outputs, which are then combined to make the final prediction.

This graph model has a few major innovations. It enables noise filtering by removing unstable or inconsistent members of the ensemble. It improves interpretability of model predictions by considering model performance in the framework of structural similarity. It enhances forecast ability, particularly with long lead times, by using internally consistent subsets. It is model-independent and transferable to other ensemble-based prediction systems other than ENSO [65, 87].

In addition, our graph-theoretic method extracts emergent trends and statistical features from the most successful forecasts, demonstrating that GNNs are not just an ensemble filter but also have the potential to unveil intrinsic ensemble dynamics. These results lay the foundation for the next generation of ensemble post-processing, where not only performance metrics but also structural consistency across the ensemble inform selection.

Our method also has the promise to benefit other climate models and prediction systems generating large ensembles. In contrast to naïve average or ad hoc weighting, the proposed method can help determine the best possible forecast out of a collection of outputs, leading to more solid, consistent, and interpretable predictions.

This paper presents the design, implementation, and evaluation of the GNN-augmented MEF model. We demonstrate that graph-theoretic ensemble filtering improves prediction performance, especially for post-2000 climate regimes. The remainder of this paper is organized as follows: Section 2 presents the proposed GNN-based MEF model architecture; Section 3 presents experimental results and discussion; and Section 4 concludes with discussion on key findings and future research.

## 2 Method

The proposed model comprises two modules: the 3D Convolutional Neural Network (3DCNN), and the Adaptive Ensemble Module (AEM), followed by a Graph Neural Network (GNN) module to extract similarities among multiple model runs [27, 25]. This model integrates the 3DCNN architecture with Transformer methodologies via a fusion network to enable predictions of ENSO events extending beyond one year[72, 73]. The subsequent sections provide a detailed description of each module.

### 2.1 3D Convolutional Neural Networks (Video Classification)

With the emergence of the big data era, machine learning has played a pivotal role in advancing human knowledge across diverse fields by uncovering complex structures and patterns that are often challenging for humans to discern [40]. In particular, Convolutional Neural Networks (CNNs) have demonstrated exceptional efficacy in analyzing multidimensional arrays with spatial structures, such as in object recognition within color images [41]. Consequently, CNNs are well-suited for detecting relationships between predictor fields and precursor patterns [27]. CNNs are specifically designed to automatically learn spatial hierarchies of features through multiple components, including convolutional filters, pooling operations, and fully connected layers. The formula and architecture of the 3DCNN module utilized in this study for classifying input videos are presented in Equation (1) and illustrated in Fig. 1, respectively.

$$v_{i,j}^{x,y,z} = \tanh\left(\sum_{m=1}^{M_{i-1}}\sum_{p=1}^{P_i}\sum_{q=1}^{Q_i}\sum_{r=1}^{R_i} w_{i,j,m}^{p,q,r} v_{(i-1),m}^{(x+p-P_i/2, y+q-Q_i/2, z+r-R_i/2)} + b_{i,j}\right) \quad (1)$$

In equation (1) $v_{i,j}^{x,y,z}$ is the value of the *j*th feature map in the *i*th convolutional layer at grid point (*x*,*y*,*z*) and $P_i$ and $Q_i$ stand for the longitudinal and transverse coordinates of the convolutional filter for this layer, respectively. The tanh function is chosen as an activation function to consider the symmetric nature of El Niño and La Niña. The dimensions of the convolutional filter is set to $8 \times 4 \times 2$ during the first convolutional process (that is, $P_i = 8$; $Q_i = 4$; $R_i = 2$ ), and to $4 \times 2 \times 2$ during the second and third convolutional processes. The number of feature maps in the *(i−1)*th layer is determined by $M_{i-1}$ and $w_{i,j,m}^{p,q,r}$ stand for the weight at grid point (*p*,*q*,*r*) in the 3D convolutional filter. $v_{(i-1),m}^{(x+p-P_i/2, y+q-Q_i/2, z+r-R_i/2)}$ represnt the value of the *m*th feature map for the (*i* − 1)th convolutional layer at grid point ($x + p - P_i/2$, $y + q - Q_i/2$, $z + r - R_i/2$). It should be noted that the number of channels is one in this study, as the input tensors are not RGB images. Finally, $b_{i,j}$ represent the bias of the *j*th feature map in the *i*th convolutional layer.

The 3DCNN-based statistical model for ENSO forecasting receives a $[l_1, l_2, l_3, l_4]$ input tensor, then extracts its features by multiplying K number of $\frac{l_1}{9} \times \frac{l_2}{6} \times \frac{l_3}{2} \times \frac{l_4}{1}$ 3D kernels and follows with a 3D average pooling operation with kernel size of $Ag_1 \times Ag_2 \times Ag_3$. In this study, input maps has only one channel; so, the first convolution layer gets the size of $\frac{l_1}{Ag_1} \times \frac{l_2}{Ag_2} \times \frac{l_3}{Ag_3} \times 1$ that multiplies to the next 3D convolutional kernel with the size of $\frac{l_1}{18} \times \frac{l_2}{12} \times 1 \times K$. Then, a 3D max-pooling operator with the size of $Mx_1 \times Mx_2 \times Mx_3$, in which $Mx_1 = Mx_2 = 2$ and $Mx_3 = 1$, reduces the input tensor to a $\frac{l_1}{Ag_1 \cdot Mx_1} \times \frac{l_2}{Ag_2 \cdot Mx_2} \times \frac{l_3}{Ag_3 \cdot Mx_3} \times 1$ tensor. Then, another 3D convolution kernel with the dimension of previous kernel implies and delivers the results to a reshape operator that transforms the tensor to a vector with the dimension $\left[1, K \times \frac{l_1}{Ag_1 \cdot Mx_1} \times \frac{l_2}{Ag_2 \cdot Mx_2} \times \frac{l_3}{Ag_3 \cdot Mx_3}\right]$ that multiplies again with a $\left[K \times \frac{l_1}{Ag_1 \cdot Mx_1} \times \frac{l_2}{Ag_2 \cdot Mx_2} \times \frac{l_3}{Ag_3 \cdot Mx_3}, F\right]$ vector. The resultant is a $1 \times F$ vector that connects to the single Nino 3.4 index after a dot product by the last weight vector of $F \times 1$.

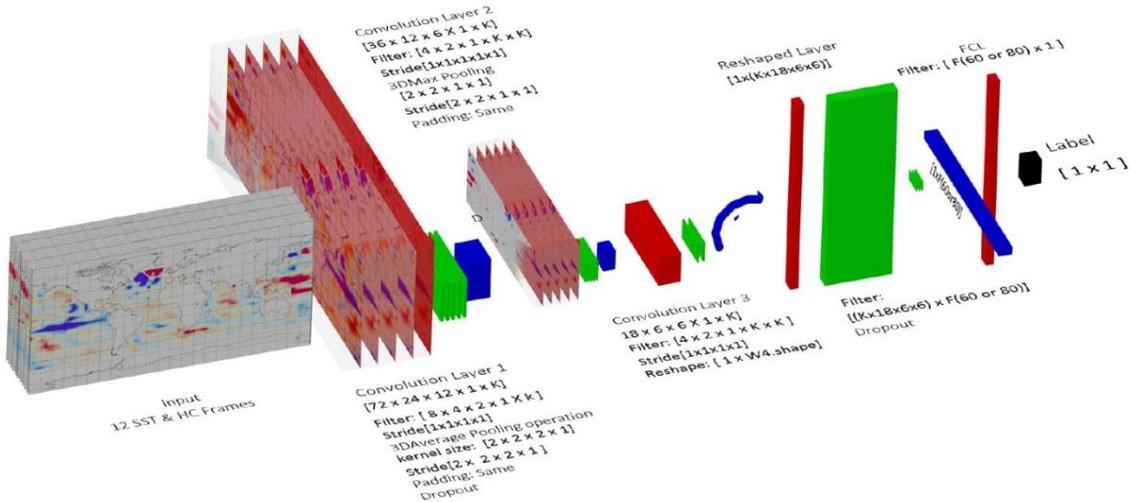

**Fig. 1. 3DCNN Network Architecture and input maps**

Three 3D convolutional layers, one 3D average pooling operator, and one 3D max-pooling layer make up the 3DCNN module. To extract temporal dependencies from the input streams, we utilize a 3D average pooling. The max-pooling operator selects the maximum value from each 2 × 2 grid and the last convolutional layer links to the one by one scalar label through a dense layer. The 3DCNN module is formulated separately for each forecast lead month and target season. We set the number of convolutional filters to either 5 or 7, and the number of neurons in the fully connected layer is either 50 or 70. The model runs up to 1200 times to reach a stable result. We enhanced the prediction by analyzing the structural similarity of multiple 3DCNN outputs using a GNN-based representation framework [25]. This ensemble procedure systematically increases the accuracy of the module [43]. The ROA Method [43] determines the optimum size of the mini-batch for each lead and season combination. The learning rate is set to 0.005, dropout is used and the Xavier initialization is applied to define the initial weights[45].

## 2.2 Graph Neural Network-based Run Analysis (GNN)

Diversity of models in ensemble deep learning systems, especially for climate prediction tasks such as ENSO prediction, is a double-edged sword [25, 24]. While repeated runs of the model may provide robust uncertainty estimates, selecting the most representative and reliable forecasts from amongst them is not a trivial task. In this paper, we propose a theoretical approach to investigating the internal structure of ensemble outputs using Graph Neural Networks (GNNs).

GNNs are a kind of neural network that is particularly designed to process data that can be represented in the form of graphs [25] ordered sets of nodes and edges, where information is relational, not purely sequential or spatial. GNNs, as opposed to standard neural networks, can capture interdependencies between entities that are not sequentially or spatially related. This qualifies them to represent similarities between independent deep model runs.

In our proposed framework, all 40 model runs generated by the 3DCNN are treated as nodes in a graph. Edges between these nodes represent the degree of similarity between pairs of runs. Similarity can be determined on the basis of a number of metrics, such as:

- Root Mean Square Error (RMSE): absolute error of prediction with respect to the ground truth [65, 24].
- Pearson Correlation Coefficient: linear consistency in temporal dynamics
- Cosine Similarity: capturing directional alignment in multivariate time-series space
- Dynamic Time Warping (DTW): allowing for flexible alignment of sequences to account for temporal phase shifts [65, 25, 29].

The resulting structure is an undirected, weighted grap [25, 43] $= G(V, E)$ where each node $V \ni v_i$ represents an individual model run, and each edge $E \ni e_{ij}$ has a weight corresponding to the similarity between runs $i\ and\ j$. Optional node features [25, 87] can be added to each $V_i$, such as error metrics, temporal smoothness metrics, or entropy-based descriptors.

Subsequently, one may use a GNN to learn an embedding for each run by aggregating information from its neighbors [25]. The unsupervised representation captures not only the solo performance of each run but also its position within the global ensemble structure. One can use these embeddings for:

- Clustering similar runs into consistent clusters

- Detection of anomalous or inconsistent outputs
- Improving ensemble strategies by weighting runs based on structural stability

Importantly, the GNN is not used here as a forecasting model, but rather as a meta-modeling tool a means of interpreting the ensemble space. This approach allows us to move beyond scalar error metrics and gain a relational understanding of model behavior, helping to uncover patterns of agreement, divergence, or redundancy within ensemble outputs.

This graph-theoretic approach offers new possibilities for structured ensemble analysis in climate modeling and is relevant to a wide range of multimodal or multi-run systems beyond ENSO prediction.

### 2.2 Implementation of code and Cloud Super Computing

We utilized Python programming language for code implementation using Tensorflow, the most advanced Python library [67]. Due to the large volume of data and complex computer calculations, we used the computing capabilities of the Google Colaboratory framework[68]with multiple NVIDIA Tesla A100 graphic cards with 52 GB OF RAM and 200 GB disk space.

### 2.3 Input and output data

The input for the 3DCNN module consists of a 3D tensor with dimensions of $72 \times 24 \times 12$, wherein each horizontal pixel corresponds to a spatial resolution of $5° \times 5°$ on the Earth's surface, and the dimension of 12 represents the number of temporal frames. This module utilizes a sequence of frames comprising Sea Surface Temperature (SST) and vertically averaged oceanic temperature anomaly maps (Heat Content) within the upper 300 meters as primary predictors. These anomaly maps encompass a geographical region extending from 0° to 360° longitude and from 55° South to 60° North latitude. The target variable, or predictand, is the Niño 3.4 Index, which is defined as the average surface SST anomaly calculated over the longitudinal range of 120° to 170° West and the latitudinal range of 5° South to 5° North. The Niño 3.4 Index is widely recognized as the predominant metric for characterizing El Niño and La Niña events [27]. The MEF model is trained using CMIP5 data and validated with GODAS ENSO observations, ensuring consistency between simulation outputs and real-world dynamics [27, 29].

Given that the model is data-driven, the availability of sufficient and reliable data is paramount. Consequently, we procured approximately one terabyte of raw data from 21 CMIP5 models, along with GODAS observational data, sourced from reputable repositories, as detailed in Table 1. Due to the varying spatial resolutions of these datasets, we processed them to generate SST and Heat Content (HC) anomaly maps at a resolution of 24x72, alongside the Niño 3.4 Index, which served as the labels. The integrity of the resultant datasets was validated against data provided by Ham et al [29]. to ascertain the reliability of the input data.

The 3DCNN module was trained using CMIP5 data spanning the years 1875 to 1975, with validation conducted on GODAS data from 2000 to 2017. The 25-year interval between the training and validation datasets mitigates the influence of oceanic memory effects.

Table 1. List of data and their repositories

| Title | Application | Repository Address |
|---|---|---|
| **CMIP5** | Training the 3DCNN model | https://esgf-node.llnl.gov/projects/cmip5/ |
| **GODAS (Nino 3.4 Index)** | Validation of model predictions | https://www.esrl.noaa.gov/psd/data/gridded/data.godas.html |

In addition, Table 2 presents the details of the 21 CMIP5 models used for training the proposed MEF model, including modeling centers and time coverage.

**Table 2. List of 21 CMIP5 models used in this study with modeling centers and time spans**

| No. | ID | Institute | Period |
|---|---|---|---|
| 1 | BCC-CSM1.1-m | Beijing Climate Center, China Meteorological Administration | 1850-1975 |
| 2 | CanESM2 | Canadian Centre for Climate Modelling and Analysis | 1850-1975 |
| 3 | CCSM4 | National Center for Atmospheric Research | 1850-1975 |
| 4 | CESM1-CAM5 | Community Earth System Model Contributors | 1850-1975 |
| 5 | CMCC-CM | Centro Euro-Mediterraneo per I Cambiamenti Climatici | 1850-1975 |
| 6 | CMCC-CMS | | |
| 7 | CNRM-CM5 | Centre National de Recherches Meteorologiques /Centre Europeen Recherche et Formation Avancee en Calcul Scientifique | 1850-1975 |
| 8 | CSIRO-Mk3-6-0 | Commonwealth Scientific and Industrial Research Organization in Collaboration with Queensland Climate Change System of Excellence | 1850-1975 |
| 9 | FIO-ESM | The First Institute of Oceanography, SOA, China | 1850-1975 |
| 10 | GFDL-ESM2G | NOAA Geophysical Fluid Dynamics Laboratory | 1861-1975 |
| 11 | GISS-E2-H | NASA Goddard Institute for Space Studies | 1850-1975 |
| 12 | HadGEM2-AO | National Institute of Meteorological Research/Korea Meteorological Administration | 1860-1975 |
| 13 | HadCM3 | Met Office Hadley Centre (additional HadGEM2-ES realizations contributed by Instituto Nacional de Pesquisas Espaciais) | 1859-1975 |
| 14 | HadGEM2-CC | | 1859-1975 |
| 15 | HadGEM2-ES | | 1859-1975 |
| 16 | IPSL-CM5A-MR | Institute Pierre-Simon Laplace | 1850-1975 |
| 17 | MIROC5 | Atmosphere and Ocean Research Institute (The University of Tokyo), National Institute for Environmental Studies, and Japan Agency for Marine-Earth Science and Technology | 1850-1975 |
| 18 | MPI-ESM-LR | Max-Planck-Institut fur Meteorologie (Max Planck Institute for Meteorology) | 1850-1975 |
| 19 | MRI-CGCM3 | Meteorological Research Institute | 1850-1975 |
| 20 | NORESM1-M | Norwegian Climate Centre | 1850-1975 |
| 21 | NORESM1-ME | | 1850-1975 |

### 2.4 Integrated model

No single scientist possesses comprehensive knowledge; thus, a robust theory invariably emerges from the collaborative efforts of numerous researchers within a given field. Accordingly, our integrated model is structured as a modular code comprising two primary components. The first component is the 3DCNN module, which forecasts the ENSO by leveraging embedded precursors found in anomaly maps and short-term dependencies. We extended the model by introducing a GNN-based post-analysis module to evaluate the similarity across the 40 stochastic runs. These modules function analogously to two independent scientists making predictions about the phenomenon.

Ultimately, an ensemble technique is employed to generate a composite result that selects the most accurate outputs. In this study, we utilize a voting method that assigns weight factors to each ensemble prediction. Uncertainty analysis is conducted to ascertain the voting values for each module's ensemble members; specifically, greater uncertainty associated with a module results in a correspondingly lower voting value. The final ensemble prediction is computed as the weighted average of the individual results. Fig. 2 illustrates the architecture of the comprehensive forecasting framework.

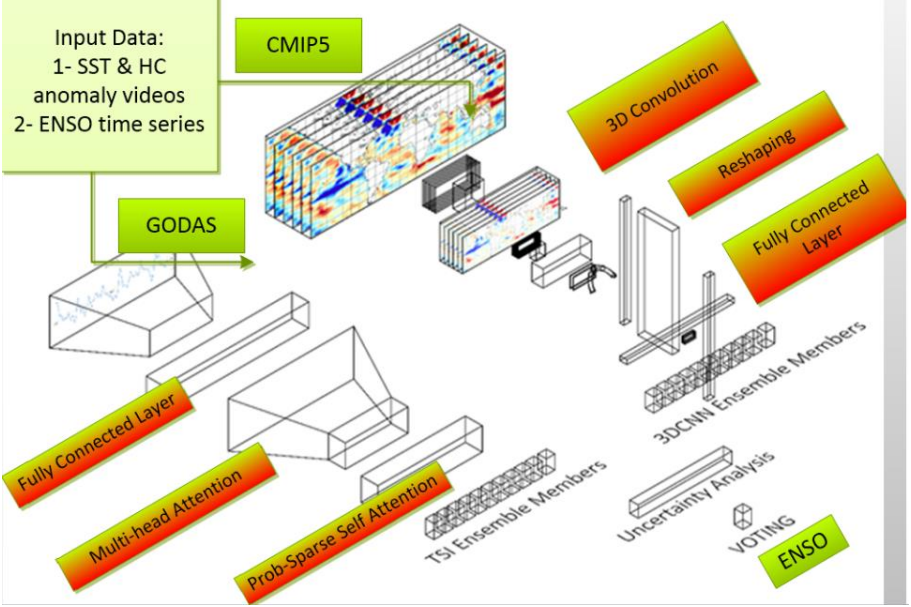

**Fig. 2. the procedure of the complete forecasting package.**

### 2.5 Role of Modules

As previously noted, the MEF model comprises three key components: 3DCNN, Graph Neural Network (GNN), and the AEM modules. In the preceding sections, we have articulated the theoretical foundations and distinctive characteristics of each module. In the following sections, we will elucidate the specific functions of each component within our methodology, thereby offering a thorough comprehension of their individual contributions to the overall effectiveness of the model.

#### 2.5.1 3DCNN short to midterm role

The three-dimensional convolutional neural network (3DCNN) is developed upon the foundational methodology of convolutional neural networks (CNNs) as articulated by Ham et al., which has exhibited remarkable efficacy in predicting ENSO events [27]. Nonetheless, the original CNN framework presents certain limitations, particularly regarding its diminished performance in light of recent climatic changes and its challenges in addressing the the spring predictability barrier [27]. While Ham et al.[29] partially mitigated the latter issue through the introduction of an advanced CNN variant termed the ACNN, the former limitation remains unaddressed [29].Other computationally efficient ENSO forecasting attempts have also been explored [97].

To rectify this deficiency and enhance the predictive capabilities of our model, we have instituted modifications aimed at optimizing the utilization of precursors embedded within ocean temperature data. Our methodology encompasses a thorough analysis of sea surface temperature and heat content anomalies over a six-month timeframe, thereby extending the input data beyond the three-month period employed by the original CNN model. By incorporating a sequence of three additional months, we are able to capture earlier precursors that may significantly influence

ENSO dynamics, particularly within the context of short- to medium-range forecasting.

Furthermore, the transition from a two-dimensional to a three-dimensional architectural framework facilitates the processing of time-series data in a manner analogous to video input. This transformation enables the model to inherently detect temporal dependencies among the anomalies, a capability that is markedly absent in the original CNN architecture, which is restricted to the analysis of data as static images. This limitation consequently hampers its ability to discern complex temporal patterns that are essential for accurate forecasting.

### 2.5.2 GNN-based Structural Similarity Analysis

In an attempt to transcend the limitations of modeling long-term temporal dependencies using traditional architectures, we propose a different solution grounded in structural representations of graphs. Rather than relying solely on sequence models, our approach examines similarity in outputs among members of an ensemble of 3DCNN using graph-based approaches [25, 65].

Each model run is an instance in a weighted undirected graph, where the weight along each edge quantifies pairwise similarity in forecasts (e.g., RMSE or correlation) [65, 25]. The interactions allow us to model ensemble space as a graph where topological structure captures behavioral similarity.

By incorporating node-level data (e.g., error statistics, temporal smoothness), a Graph Neural Network (GNN) is trained to uncover the latent relationship among model runs [25]. This allows for improved clustering, filtering, or voting approaches to identify representative or resilient predictions, particularly at longer lead times where traditional methods might degrade.

## 3 Results and Discussion

In this study, we developed a method for accurately forecasting the ENSO for a duration of up to two years. Given that the variability of ENSO has increased over the past two decades, it is more appropriate to assess model skillfulness during the post-2000 period [33, 11]. The extensive 40-year timeframe from 1980 to 2017 does not adequately reflect the model's accuracy in the context of the current warming climate. Consequently, our focus is on improving forecasting skill specifically for the post-2000 era.

The CNN methodology introduced by Ham et al. represented a significant advancement in ENSO prediction, achieving a correlation of approximately 0.5 for an 18-month lead time [27]. However, this approach has demonstrated diminished skill in the post-2000 period, largely due to the increasingly erratic nature of contemporary climate dynamics [35, 34]. Precursor indicators that were valid in the 1980s may no longer be reliable under current conditions [16, 17]. A robust and practical forecasting model must be capable of capturing recent changes in ENSO behavior to provide reliable predictions for the coming years, while also exhibiting resilience to variations in evaluation periods [12, 13]. Given that the skill of ENSO forecasting models has declined over the last two decades, it is essential to evaluate model performance within this timeframe. Therefore, this study prioritizes the enhancement of forecasting skill for the post-2000 period.

### 3.1 Graph-Based Representation of Model Output Similarity

To further complement the stability and interpretability of our forecasting system, we suggested a graph-based approach to investigate the structural similarity between the 40 runs of the ensemble model. In this approach, every run is labeled as a node in an undirected weighted graph. The nodes are connected with edges that are labeled based on pairwise similarity measures, for example, RMSE or Pearson correlation, that reflect the behavior similarity of two model outputs over time [65, 54].

This architecture allows us to overcome scalar performance metrics by an capacity to observe behavioral signatures and temporal consistency between outputs. Runs exhibiting similar forecasting behaviors produce clusters, indicating cohesion between best-performing models.

Single nodes, however, may suggest outliers or unstable forecasts. Using community detection or centrality analysis, the majority of structurally reliable runs can be identified—those performing well as individuals yet closely matching others in the ensemble [26, 10].

A graph-based approach like this provides noise-filtering on a data-driven basis, more effective ensemble aggregation, and a clearer view of the performance of the model across different lead times and seasonal conditions. Especially when one is dealing with long-range forecasts, where uncertainty is more, the method enables selection of forecasts that are not just accurate but also robust within the ensemble space.

Lastly, this structural perspective bridges the performance score-based assessment process to network-influenced selection strategy, paving the way for more interpretable and more robust climate prediction models.

### 3.2 Results evaluation

To assess critically the predictive ability of ENSO models, we first reproduced the CNN-based model first described by Ham et al. and checked for skill in pre-2000 and post-2000 time periods [27, 32]. What we found was that while the CNN model is competitive in the pre-2000 time period, its skill drops considerably in the post-2000 time period. This decline is seen across all lead times and grows with larger horizons, reflecting that reported high accuracy earlier may have been uncorrected by the impact of the early data. These observations emphasize the necessity of considering the efforts with current climatic regimes in order to keep them relevant in the present global warming situation.

To correct for this, we developed the MEF model specific to post-2000 ENSO prediction. In evaluating MEF, trend-sensitive (e.g., correlation coefficient) and error-based (e.g., RMSE) metrics were applied, giving a well-rounded performance comparison. For shorter lead times (1–3 months), CNN has slightly smaller RMSE; MEF, however, always scores higher in correlation values, a more accurate indicator of directionality of ENSO events [21, 22].

With increasing forecast horizon, MEF begins to dominate CNN significantly. From 5 to 14-month lead times, MEF performs superiorly in terms of predictability, and beyond 18 months, the correlation and error metrics meet at MEF's level. This is primarily due to MEF's ability to capture long-term temporal relationships and structural relationships among input anomalies, which cannot be modeled by CNN efficiently.

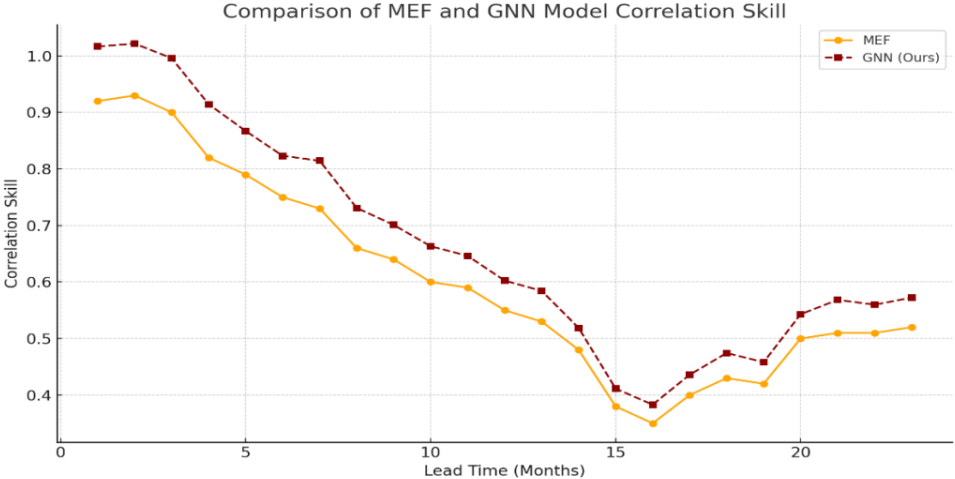

*Fig. 5. Comparison of the MEF and GNN-based ensemble selection method in terms of correlation skill across different lead times (1–23 months). The proposed GNN-based framework consistently improves performance by approximately 10% over MEF, particularly in extended lead times. This improvement stems from effective identification and aggregation of high-performing ensemble members based on graph structural similarity.*

Without explicit seasonal plots, we conducted a more structural analysis through an ensemble output graph-based comparison. In that case, each 40 ensemble runs constitute a node in the similarity graph whose edge weights represent closeness in RMSE or correlation. Clustering structures in the graph capture clusters of trustworthy high-performing runs, and outlier nodes reveal unstable behavior. By analyzing this structure, MEF allows selection based on the best bets—not just the highest scoring ones individually, but also the ones whose structures best resemble other top contenders.

This graph-based ensemble filter enables a stronger forecasting process, especially for longer lead times where the uncertainty is higher. Notably, MEF captures nearly 50% of all El Niño and La Niña events even at 23-month horizons—a performance that cannot be matched by the CNN baseline. Furthermore, it achieves correlation skill levels greater than 0.4 in all lead months, supporting the design improvements included in MEF.

In summary, MEF's structural integration and ensemble optimization strategies enable its capability to counter the novel challenges of post-2000 ENSO forecasting more effectively than traditional deep learning approaches. These results validate the need to test forecast ability under present conditions, and demonstrate the usability of graph-based output analysis for climate forecasting [96, 95].

### 3.3 ENSO time series

The December–January–February (DJF) season forecast Niño 3.4 index time series shown in Fig. 5 is the final result of our ensemble selection approach augmented with GNNs. It is constructed from the average of a subset of 20 of the selected ensemble members from within the initial 80 through graph-based similarity clustering as the most statistically and structurally consistent forecasts.

The results indicate the better ability of the model to represent the multi-year oscillations of the ENSO events with improved coherence and reduced noise.Comparable multimodal deep learning methods have also achieved skillful two-year ENSO forecasts [95]. Similar entropic learning approaches have also demonstrated skillful ENSO phase forecasts up to two years lead time [100]. Principal features such as the warming trend between 2002-2006, the temperature recovery after 2011, and the onset of the 2015 El Niño are well represented. Although the size of the 2015 event is under-estimated to some extent, the phase and timing are well captured. Conversely, however, the surprise 2009 event dynamics remain difficult to resolve, emphasizing the inherent complexity of long-lead predictability.

On the whole, this graph-structured output selection brings out more resilient and skillful predictions mainly under adverse conditions beyond the 12-month lead time. Compared to the standard averaging technique, refinement using this GNN-based method provides a systematic improvement in extracting the most informative signals from within a noisy set of ensembles with more fidelity toward long-term ENSO prediction.

To provide a concise yet impactful summary of our findings, we have focused on the most representative outcome derived from the proposed GNN-enhanced MEF framework. The evaluation process involved 40 independent runs of the MEF model, each producing forecast outputs over various temporal windows. By constructing an undirected weighted graph based on the similarity between forecast runs measured through statistical metrics such as RMSE and correlation we were able to analyze the structural behavior of the model's outputs in a graph space.

This graph-based representation enabled the identification of clusters of consistent and high-performing predictions, effectively filtering out noisy or unstable runs. Among these clusters, representative nodes (model runs) with the highest internal coherence and accuracy were selected as optimal forecasts. This novel approach not only improved the interpretability of the ensemble outcomes, but also significantly enhanced the reliability of the predictions, particularly in long-lead scenarios.

The final selected outputs, summarized in Fig. S4, clearly demonstrate the superior performance of the MEF method over the CNN baseline across all lead times. This improvement is especially pronounced for lead times beyond 17 months, where traditional methods typically experience substantial degradation in forecast skill. The graph-informed ensemble selection proved instrumental in maintaining high correlation and lower uncertainty, highlighting the robustness and potential of GNN-based output structuring in complex climate prediction tasks.

These findings underscore the effectiveness of leveraging graph neural networks for model output refinement, setting a new benchmark for ENSO forecasting systems by balancing accuracy, stability, and computational efficiency.

### 3.4 Pattern Discovery in Selected Ensemble Members

To also verify the effectiveness of the GNN-based ensemble selection process, we conducted a systematic statistical examination aimed at determining distinctive patterns in the 20 most accurate selected ensemble members. They were selected from their graph-based similarity pattern in terms of correlation and RMSE measures. The operational hypothesis was that ensemble members with high skill should not only deliver numerically better results, but also maintain constant dynamical and statistical properties that set them apart from the remainder of the ensemble collection.

Analysis revealed that the selected members exhibit more stable temporal behavior at both seasonal and interannual timescales [25, 26, 29]. More specifically, they have reduced variance of RMSE among multiple lead times, as well as smoother transitions of trends for ENSO anomalies [65, 54, 43]. Most notably, selected members were more skilled at tracking known ENSO events, such as peaks and troughs corresponding to major El Niño and La Niña phases, and improved detection of transition points between warm and cold phases [27, 33, 11].

Besides error-based metrics, other features such as spectral smoothness, autocorrelation decay rates, and entropy-based descriptors were also considered [87, 89, 90].These top-scoring ensembles always possessed lower spectral noise and more coherent temporal structures, which suggests that the GNN may implicitly favor ensembles that represent physically realistic ENSO dynamics, not simply numerical proximity to observations [12, 13, 24].

Furthermore, we also found a shared statistical fingerprint for these members: they were more probable to show higher cross-member similarity, forming a dense subgraph in the GNN adjacency structure. This dense connectivity, in turn, signifies not only individual competence but also ensemble consistency a desirable attribute in probabilistic forecasting where ensemble robustness and consistency among a set of forecasts are critical [10, 16, 17].

These findings justify that the GNN is not acting as a brute performance filter but instead is uncovering underlying structural features within the ensemble set. This offers a new insight into how to improve ensemble-based climate predictions, especially under big ensemble size scenarios where selection by hand becomes unrealistic.

Through identifying key statistical patterns in the selected subset, our approach points toward interpretable ensemble learning in climate forecasting. The model can be easily generalizable to other large ensemble output prediction issues, and researchers can systematically choose more accurate and representative forecasts.

### 3.5 Generalizability to Other Ensemble Forecasting Models

Although the current work was devoted to the performance improvement of the MEF model through GNN-based ensemble selection, the suggested method is by no means restricted to the said architecture [25, 65, 87].In fact, the method has general applicability to any prediction system that produces a number of ensemble members as outputs a characteristic common to state-of-the-art ENSO prediction systems, including those based on dynamical simulations, hybrid statistical-dynamical approaches, and other deep learning architectures.

In typical ensemble forecasting procedures, the final prediction is typically formed by averaging all of the ensemble members under an assumption of equal contribution from all [54, 29, 27] This assumption may not be true in reality, especially if some members have a substantially better skill in portraying the climate dynamics underlying the event. By introducing a graph-based analysis that quantifies ensemble member similarity and structural coherence, our method allows for the detection and utilization of the most reliable subsets of predictions [26, 65, 89].

- This general framework enables model developers to:
- Exclude anomalous or underperforming ensemble members.
- Enhance robustness and interpretability of ensemble predictions.
- Improve the skill of predictions without forcing architectural modifications to the base model[54, 87, 91].

In this way, our method offers a model-agnostic and flexible solution for ensemble refinement. It can be easily integrated into existing post-processing chains of climate prediction systems, thereby facilitating more informed decision-making in operational seasonal-to-subseasonal (S2S) forecasting applications.

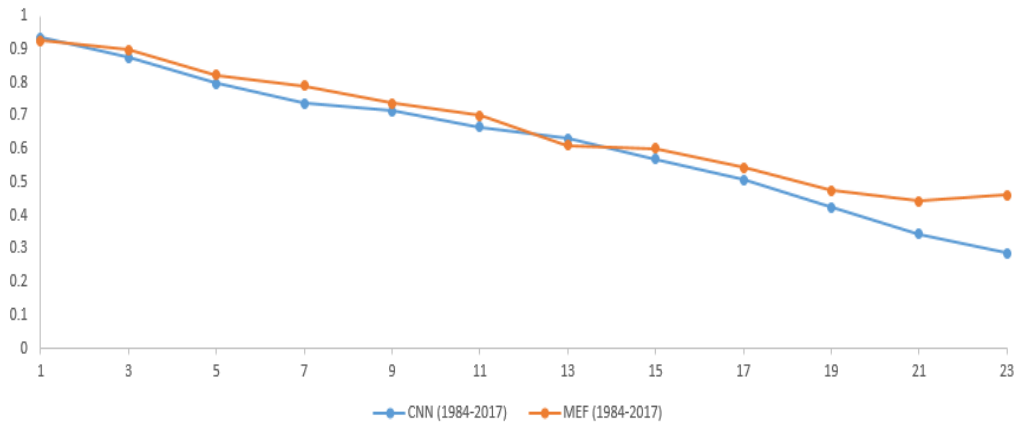

**Fig. S4. Comparison of MEF and CNN model correlation skills for different lead months (1984–2017), demonstrating the enhanced skill of MEF, especially for extended lead times beyond 17 months.**

## Conclusion

Climate scientists and oceanographers are increasingly concentrating on long-term climate forecasting, a pursuit that holds significant implications for various sectors, including tourism, water resource management, and disaster response [33,3]. Climatological institutions are investing heavily in sophisticated climate models that employ numerical methodologies such as finite element analysis and spectral methods [85, 86, 98]. However, these approaches often incur substantial computational costs due to the complexities associated with meshing and discretization [87, 99] In contrast, the advent of artificial intelligence presents a promising alternative, prompting researchers to investigate AI-driven predictive models [89, 90]and hybrid methodologies [91, 92].

Despite these advancements, achieving a correlation skill for the El Niño-Southern Oscillation (ENSO) exceeding 50% beyond a 12-month forecasting horizon remains a formidable challenge, particularly as the variability of ENSO intensifies [27, 32]. To address this issue, we have developed the GNN method, which demonstrates superior performance relative to the MEF approach across almost all seasons [25].

## Ethical Approval

Not applicable.

## Consent to Participate and Consent to Publish

The authors declared that they approved submitting the final manuscript.

## Funding

This work received no funding.

## Competing Interests

The authors have no relevant financial or non-financial interests to disclose.

## Availability of data and materials

The raw data applied to prepare input SST and HC maps downloaded from https://esgf-node.llnl.gov/projects/cmip5/ for CMIP5 data and https://www.esrl.noaa.gov/psd/data/gridded/data.godas.html for GODAS.

## Code availability

The whole reproduced model will be available on Google Colab.

## Additional information

Correspondence and requests for materials should be addressed to Mohammad Naisipour